\documentclass{article}

\usepackage{arxiv}

\usepackage[utf8]{inputenc} 
\usepackage[T1]{fontenc}    
\usepackage{hyperref}       
\usepackage{url}            
\usepackage{booktabs}       
\usepackage{amsfonts}       
\usepackage{nicefrac}       
\usepackage{microtype}      
\usepackage{graphicx}

\title{{\it Ab initio} electronic stopping power for protons in Ga$_{0.5}$In$_{0.5}$P/GaAs/Ge triple-junction solar cells for space applications}

\author{
  Natalia E.~Koval\\
  CIC Nanogune BRTA\\
  20018 Donostia-San Sebasti\'an, Spain\\
  \texttt{koval.tasha@gmail.com} \\
   \And
 Fabiana Da Pieve \\
Royal Belgian Institute for Space Aeronomy\\ BIRA-IASB, 1180 Brussels, Belgium\\
  \And
  Emilio Artacho\\
  CIC Nanogune BRTA, 20018 Donostia-San Sebasti\'an, Spain\\
  Donostia International Physics Center DIPC, 20018 Donostia-San Sebasti\'an, Spain \\
Theory of Condensed Matter, Cavendish Laboratory, University of Cambridge, Cambridge CB3 0HE, United Kingdom \\
Ikerbasque, Basque Foundation for Science, 48011 Bilbao, Spain\\}

\begin{document}
\maketitle

\begin{abstract}
Motivated by the radiation damage of solar panels in space, 
firstly, the results of Monte Carlo particle transport simulations
are presented for proton impact on triple-junction Ga$_{0.5}$In$_{0.5}$P/GaAs/Ge 
solar cells, showing the proton projectile penetration in the cells as a
function of energy.
 It is followed by a systematic {\it ab initio} investigation of the electronic 
stopping power for protons in different layers of the cell at the relevant
velocities via
real-time time-dependent density functional theory (RT-TDDFT) 
calculations. 
The electronic stopping power is found to depend significantly on different 
channeling conditions, which should affect the low velocity damage predictions, 
and which are understood in terms of impact parameter and 
electron density along the path. 
Additionally, we explore the effect of the interface between the 
layers of the multilayer structure on the energy loss of a proton, 
along with the effect of strain in the lattice-matched solar cell. 
Both effects are found to be small compared with the main bulk effect.
The interface energy loss has been found to increase with decreasing 
proton velocity, and in one case, there is an effective interface energy gain.
\end{abstract}


\section{Introduction}

\begin{figure}[ht]
\begin{center}
\includegraphics[width=0.7\textwidth]{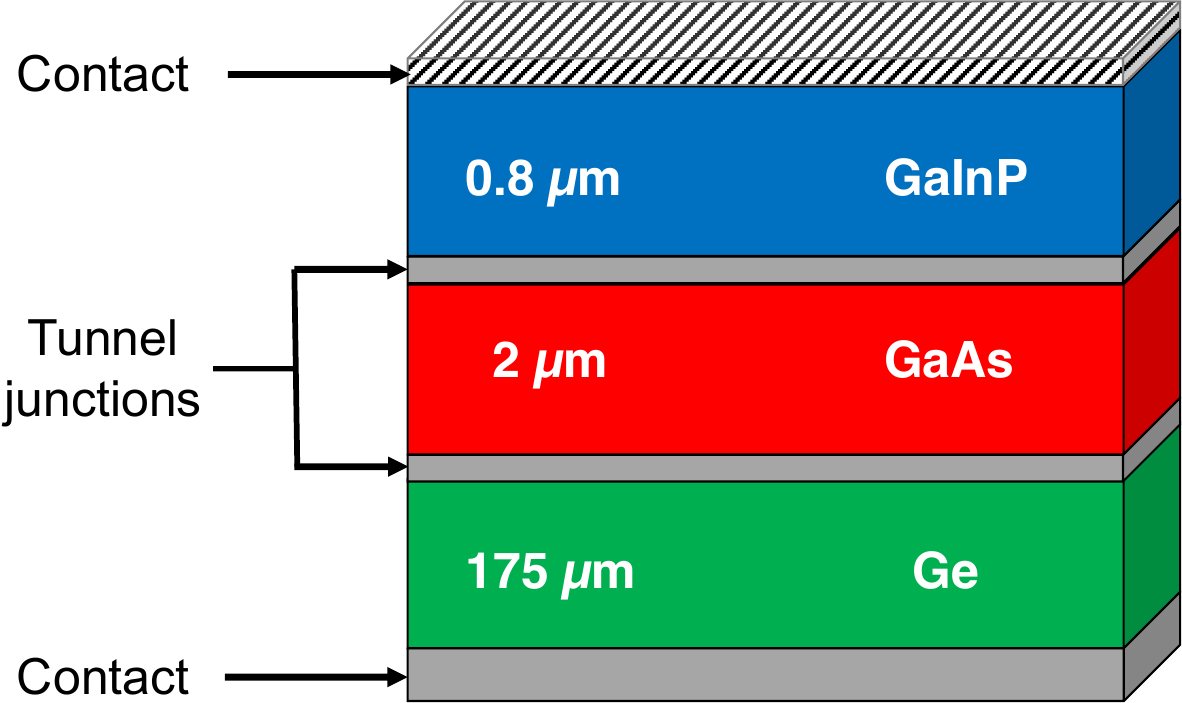}
\end{center}
\caption{\label{sc} Schematic structure of the triple-junction solar cell. The active layers (GaAs and Ga$_{0.5}$In$_{0.5}$P) are lattice matched to the Ge substrate. The tunneling junctions are usually made of p- and n- doped GaAs. Metallic contacts are often aluminum. A detailed structure can be found in Ref~\cite{tj-ss}.}
\end{figure}

With the new advancements in solar cell technology and the
rapid expansion of space missions, understanding the key aspects
which link the macroscopic response of solar cells
of current and future spacecraft to the fundamental processes of particle
stopping inside the target has acquired new relevance. 
Solar cells are key components of spacecraft and satellites as they provide
power for navigation, communication, data handling, thermal control, and
functioning of the instrumentation.
At present, the so-called triple-junction (TJ) solar cells, made of
Ga$_{0.5}$In$_{0.5}$P, GaAs, and Ge layers (see scheme in figure~\ref{sc}),
are the state of the art for space applications, generally protected by an
amorphous SiO$_2$ coverglass. All layers are usually lattice-matched to the
substrate (Ge).

In space, radiation degrades the performance of the solar cells 
mainly by cumulative effects involving atomic displacements.
Such displacements introduce consequent defect levels in the electronic 
structure, which in turn affect the output current. 
The atomic displacements are caused by the scattering of both primary incoming charges (protons) and secondary projectiles (displaced target atoms, protons, neutrons) generated by the primary radiation.
For particle energies below 10-30 MeV (depending on the material) the damaging
interaction is mainly due to the Coulomb repulsion between the proton and the target without affecting the nuclei, while hadronic interactions, 
involving sub-nuclear particles, take over at energies
above 30-50 MeV.
The Coulomb contribution to the non-ionizing energy loss 
(NIEL, the rate of energy deposition to atomic displacements, 
linked to the nuclear stopping) is the main one in the relevant
ground testing
energy range (0-10 MeV), using unidirectional fluxes on unshielded
solar cells. Such energy range and configuration have been shown to
be representative of the cumulative damage caused by omnidirectional
space radiation on shielded solar cells, where energies in general
go up to hundreds of MeV~\cite{doi:10.1063/1.119660,mess20019,summ19942068}. 

The (Coulomb) NIEL is often extracted from SRIM (The Stopping and Range of Ions in Matter)~\cite{srim,srim-book}, from
tabulated-data-based approaches like SR-NIEL~\cite{SRNIEL}, or from Monte 
Carlo particle transport codes such as
Geant4~\cite{AGOSTINELLI2003250,1610988,ALLISON2016186} 
(the latter two also provide the hadronic contribution). 
Despite some differences, some basic assumptions are similar in 
these approaches: crystalline order is not considered for the target (thus, there is 
no distinction between channeling and off-channeling conditions), 
cascades are treated within the binary-collision approximation (BCA), and the description of electronic stopping is 
based on the (perturbative) Bethe-Bloch theory at high energy and by 
the Lindhard dielectric theory combined with the local-density 
approximation (LDA) at lower energies (E/n $<$ 2 MeV), 
implying that the electronic stopping power (ESP) at each point in the
system is considered to be the same as that of a uniform electron 
gas of the same density. 
  An ad-hoc parametrization of Lindhard results is used at 
intermediate energies. In general, the low energy physics in both SRIM and Monte Carlo particle transport codes, such as Geant4, strongly relies on phenomenology and the use of experimental data.

  The influence of the ESP on defect production 
by irradiation of different ions at different energies is under intense
investigation~\cite{TOUL201183,THOM2013102,SALL2013102,WEB201519,ZARK201426,
ZARK201527,BACK201245,WEB20122,ZHA2014327,WEB20155,ZARK2015scri,
PhysRevB.99.174301,PhysRevB.99.174302,lee2020multiscale}.
  Recent studies, which address the ESP problem 
from the {\it ab initio} point of view and combine it with the molecular 
dynamics simulations of damage cascades, show the importance of including 
electronic stopping effects in 
cascade formation, in terms of the number of formed defects and cascade
morphology~\cite{PhysRevB.99.174301,PhysRevB.99.174302,lee2020multiscale}. 
  Electronic excitation, as a result of the moving primary and 
secondary projectiles in the first stages of the radiation damage process, 
can significantly alter the subsequent ionic displacements and the number 
of defects formed in the solid. 
  Thus, accurate description of the ESP processes is essential for the correct estimation 
of the radiation effect in materials.

  Real-time time-dependent density-functional theory (RT-TDDFT) has opened 
the way for the predictive first-principles description of electronic stopping 
processes in condensed-matter systems, in 
a non-linear and non-perturbative way~\cite{PhysRevLett.99.235501,
PhysRevLett.99.016104,PhysRevLett.108.213201,PhysRevB.85.235435,
PhysRevLett.108.225504,PhysRevB.89.035120,Lee2018}. 
  Recent RT-TDDFT studies on channeling conditions for proton 
impact have found a strong sensitivity of the ESP 
on crystal direction and impact 
parameter~\cite{PhysRevB.91.125203}.
  Deviations from velocity proportionality at very low 
velocity have also been obtained~\cite{PhysRevB.91.125203,KAI2017},
consistent with the finite-velocity threshold observed in 
experiments\cite{PhysRevLett.103.113201}. 
  Others have highlighted the different contributions of semicore
electrons in channeling versus off-channeling conditions~\cite{SCHL2015}.
  Local enhancement and reduction of the ESP for specific 
channels and velocity-dependent deviations from an ideal channeling 
trajectory were also reported~\cite{Lee2018}, as well as the influence
of a strongly anisotropic crystal structure on the stopping 
power~\cite{JESS2019}.

The explicit consideration of channeling trajectories in the 
computer simulation of the radiation damage has been shown to
be important~\cite{CHRISTIE2015105}.
  Both experiments and simulations show that ion channeling plays a 
significant role in radiation damage cascades in crystalline materials. 
It has been observed that for an ion moving with energies in the keV range, the penetration depth is 
much larger due to the channeling effect, which affects the spatial 
distribution and the shape of the displacement 
cascade~\cite{PhysRevLett.12.363,4331054,Race_2010}. 
  The channeling effect also contributes to the formation of the
sub-cascades~\cite{doi:10.1080/10420159008213055,STOLLER2012293,Race_2010} 
and increases the time of the defect formation~\cite{doi:10.1063/1.4922457}.

An initial Monte Carlo particle transport study (Geant4-based) shows 
that in a realistic space mission, the energy range of protons traversing
all the three layers (deriving from both the primary space radiation and 
as products of nuclear fragmentation in the layers) extends until low 
energies.
Low energy protons (up to few hundred keV), which stop inside the cell, have been shown to cause the most severe damage on TJ solar cells~\cite{Sumita2003}. For all the reasons listed above, we chose the low energy regime as the focus of this paper.

In this context, the aim of this work is to present a systematic 
investigation of the ESP of the 
sub-junction materials Ga$_{0.5}$In$_{0.5}$P, GaAs, and Ge 
of space solar cells under low-energy (keV range) proton impact in channeling conditions,
using RT-TDDFT as implemented 
in  SIESTA~\cite{Soler2002,PhysRevB.91.125203}. We consider various  channeling trajectories for the projectile (different impact 
parameters) inside the target.
  The results are compared with SRIM~\cite{srim} data, which do not discriminate channelling trajectories from others.
 
   Finally, an analysis is performed of the electronic energy loss at 
the interface between two upper layers of the TJ structure. 
  To the best of our knowledge, no study has been presented 
investigating the change of the electronic response to particle radiation across the stack. 
  The results presented here should contribute to understand 
to what extent channeling effects can be linked to both the electronic
density and the atomic number of the element(s) in the compounds 
constituting the solar cell stack, which should serve as a basis to a 
further study of non-adiabatic MD simulations of cascades. 
 
  The key questions we ask in this work are: (I) how does low-energy 
channelling electronic stopping changes with respect to the widely used
random-trajectory average, as used in SRIM; (II) how large can
the differences among channels be; (III) how different are 
the three materials; (IV) are there significant electronic
effects at interfaces; and (V) what is the effect of the
strain in the layers due to epitaxial growth.

\section{Method}

\subsection{Monte Carlo particle transport simulations through the solar cell}\label{sec:met-MC}

  1D-Monte Carlo proton transport calculations of the isotropic protons through the multilayer stack 
of the TJ solar cell have been performed with the MUlti-LAyered Shielding 
SImulation Software (MULASSIS)~\cite{lei2002} (based on
Geant4~\cite{AGOSTINELLI2003250,1610988,ALLISON2016186}, available 
via the European Space Agency's (ESA) online interface SPENVIS~\cite{heynd2003}. 
Geant4 is a Monte Carlo particle transport simulation software used in application to
high energy, nuclear and accelerator physics, as well as
studies in medical and space science. 
In order to simulate different energy ranges, different systems and configurations, a pre-defined set of physics models (the so-called physics lists) is included in the Geant4 toolkit, which allow for the description of hadronic and electromagnetic interactions with different accuracy.
The hadron/ion interactions (in general for protons, neutrons, and pions) at energies above 20 GeV, are described by a Quark Gluon String (QGS) model~\cite{ivan2012}.
At energies below 10 GeV, the Binary Intra-nuclear cascade (BIC) model for primary proton and neutron interactions with nuclei is used~\cite{ivan2012}.
At low and intermediate energies, ionization and elastic scattering are the main processes of the interaction of protons with matter.

The energy loss due to electromagnetic interactions is calculated using the energy loss, range, and inverse range tables pre-computed at initialization of Geant4 for each material~\cite{ivan2017,aposto}.
In all Geant4 electromagnetic physics configurations, two models are used for proton ionisation~\cite{bagulya}: the PSTAR/SRIM stopping power~\cite{pstar,srim} for proton energies below 2 MeV, and the Bethe-Bloch formula with shell, Barkas and Bloch, and density effect corrections~\cite{refstopG4} for energies above 2 MeV. The G4EmStandardPhysics$_{-}$option3 was used for the simulation
of electromagnetic interactions. By choosing this physics list, the
appropriate descriptions of the electromagnetic processes are automatically
applied for different particles in different energy ranges.

A simplified solar cell structure with a typical SiO$_2$ amorphous 
100 $\mu$m thick coverglass~\cite{Mess2006}, and with the Ga$_{0.5}$In$_{0.5}$P, GaAs, 
and Ge layers having 0.8, 2, and 175 $\mu$m thickness respectively, 
is used, similarly to \cite{AIERKEN201836}.
  The omnidirectional fluxes of trapped protons, covering the energy range of 0.1 to 400 MeV (Earth's radiation belt, on the basis of data from more than 20 satellites), 
have been generated via the NASA AP-8 model~\cite{vette1991}.

The Transport of Ions in Matter (TRIM) part of the SRIM code is used to calculate the proton track trajectories through the solar cell stack. TRIM is a Monte Carlo code that calculates the interactions of energetic ions with amorphous targets using several approximations, such as binary collisions only, an analytical formula for the ion-ion interactions, and the so-called concept of a Free-Flight-Path between the collisions, such that only significant collisions are evaluated
~\cite{srim,srim-book}.

\subsection{Electronic stopping power from RT-TDDFT}\label{sec:comp-set}

  The ground state configuration of each system is calculated using the
DFT implementation of the SIESTA code (see Appendix A. for details)~\cite{Soler2002} with the 
projectile placed at the initial position for each trajectory (specified in section~\ref{sec:res-stop-SRIM}). 

  The ground state Kohn-Sham orbitals of the system are 
obtained by solving the Kohn-Sham equations
self-consistently until the total ground state energy is converged.
  The exchange-correlation functional uses the local-density 
approximation (LDA) in the Ceperley-Alder form~\cite{PhysRevLett.45.566}.
 Norm-conserving Troullier-Martins~\cite{PhysRevB.43.1993} 
relativistic pseudopotentials are used to replace the core electrons (see Appendix A. for the parameters used to generate pseudopotentials).
It is known that core electrons can have large effects on
the ESP (see e.g. Ref.~\cite{PhysRevLett.121.116401}).
  However, here we only test the effect of core electrons for one proton trajectory in Ge, for all the others we consider the interaction of the projectile 
with valence electrons, given the low charge of the projectile, and the relatively low velocity range.
  In such limit the effect of core electrons is negligible as we demonstrate in our work (see 
also the validation with experiments of similar calculations for
protons in Ge in Ref.~\cite{PhysRevB.91.125203}).
  The valence electrons are represented by a double-$\zeta$ polarized 
basis set of numerical atomic orbitals defined as specified in the Appendix A.

  We use real-time time-dependent DFT (RT-TDDFT) implementation 
of the SIESTA method~\cite{PhysRevB.66.235416,PhysRevB.91.125203} to 
evolve the electronic orbitals in time. 
The time-dependent Kohn-Sham equations are solved by 
real-time propagation for discretized time, using the Crank-Nicholson
scheme~\cite{crank_nicolson_1947,PhysRevB.66.235416}
with a time
step $dt=1$ attosecond. 
  The effect of the moving basis set is accounted for by a L\"owdin
transformation, as described in Refs.~\cite{TomfohrSankey,PhysRevB.95.115155},
which is known to be adequate for relatively low velocity of the moving
basis orbitals, and offers strictly unitary propagation for finite $dt$.

  The focus of this work is on the purely ESP for constant velocity projectiles.
  The forces on the nuclei of the target atoms and on the projectile 
itself are therefore disregarded for the time propagation, thereby
describing nuclear dynamics with frozen host nuclei and a constant
velocity projectile, as done in many similar studies~\cite{PhysRevLett.99.235501,
PhysRevLett.108.225504, PhysRevB.91.125203, PhysRevLett.121.116401}.
  It allows the clean separation of the electronic and nuclear 
contributions to the total stopping.
  Fixing the atomic positions has a negligible effect for the simulations
performed in this work, given the fact that the projectile
traverses the simulation box in about $t \leq 4$ fs, and the ionic displacements on that
time scale are negligible (head-on collisions are avoided in
channelling conditions).

  All three target materials have a diamond cubic structure with a 
similar value of the lattice constant.
  We used the theoretical lattice constant of 5.59 {\AA} for Ge 
(for which the experimental value is 5.66 {\AA}) in order to compare 
our results with Ullah \emph{et al.}~\cite{PhysRevB.91.125203}. 
  For other layers, we used experimental lattice constants of 5.65 {\AA}
 for GaAs, and 5.66 {\AA} for Ga$_{0.5}$In$_{0.5}$P (average between 
the values for GaP, 5.45 {\AA}, and InP, 5.87 {\AA}). 
  The change in the lattice constant by $1.7\%$ only changes the 
stopping power by a negligible $0.28\%$ according to our 
results (see section~\ref{sec:res-strain}).

\begin{figure}[ht]
\begin{center}
\includegraphics[width=0.95\textwidth]{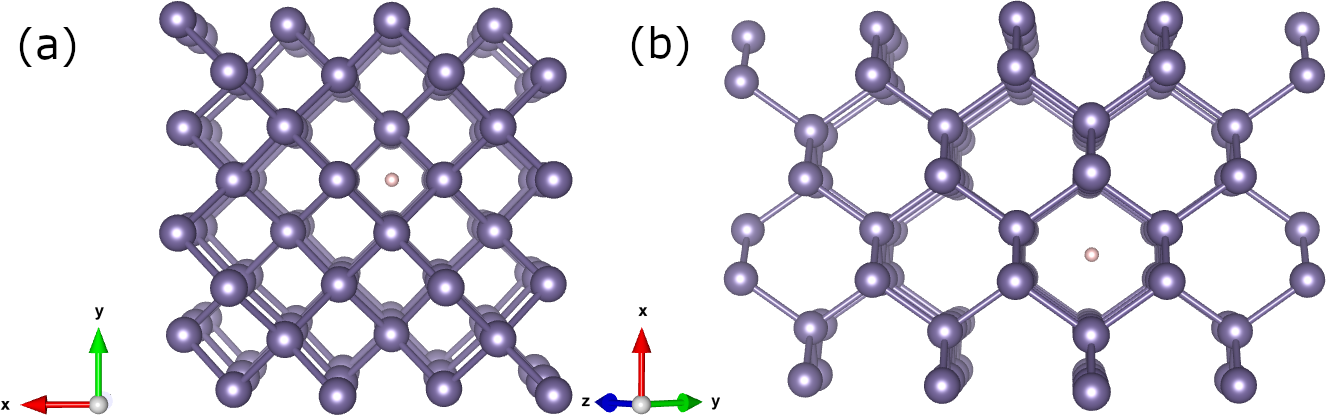}
\end{center}
\caption{\label{Figstr-Ge}  Ge super-cell (96 atoms). 
A proton position is shown in the center of (a) [001] channel and 
(b) [011] channel. GaAs and Ga$_{0.5}$In$_{0.5}$P have a similar structure.}
\end{figure}

The targets are modeled by a super-cell of 96 atoms, constructed by 
multiplying the conventional unit cell (which 
consists of 8 atoms) by $2\times2\times3$ in the $x$, $y$, and $z$ directions,
respectively. 
  A $2\times2\times1$ Monkhorst-Pack~\cite{PhysRevB.13.5188} {\bf k}-point mesh is used, which
corresponds to a {\bf k}-grid inverse cutoff of 8.4 {\AA}. The {\bf k}-points
are displaced to the centers of the grid cells.
  Two channels in the target super-cell are chosen as trajectories for
the proton, along the [001] and [011] crystallographic directions, 
as shown in figure~\ref{Figstr-Ge} for the case of Ge.  
  Besides the channeling trajectories, and defining the channel center
as the axis furthest away from all atoms, 
we also choose various off-center parallel channel trajectories with different 
impact parameters, i.e., at different closest distances of the
proton from the host atoms. The same super-cell is used in simulations for [001] and [011] channels, since the length of the trajectory is similar in both directions.
   
\begin{figure}[ht]
\begin{center}
\includegraphics[width=0.8\textwidth]{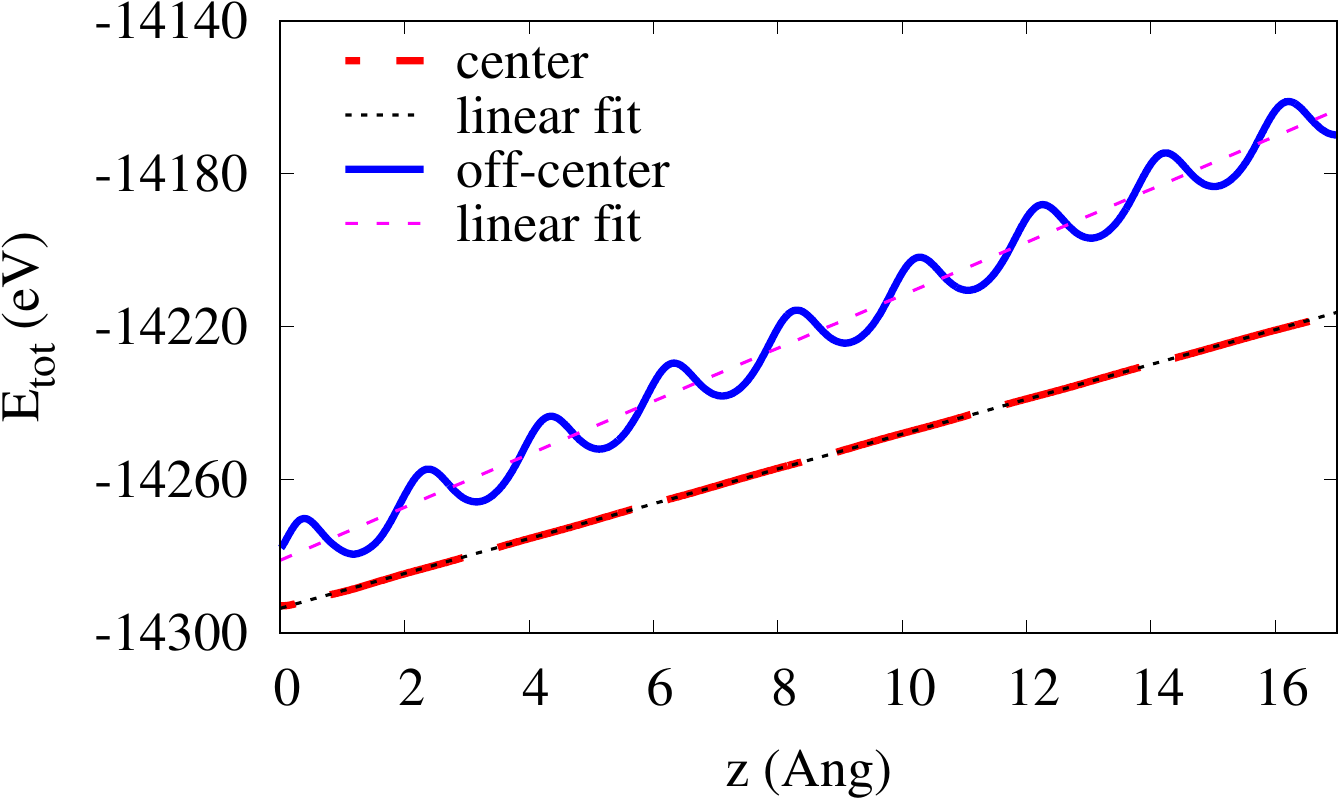}
\end{center}
\caption{\label{fitting1} 
 Position-dependent total electronic energy for two trajectories of the proton moving through Ge with velocity $v = 0.5$ a.u. 
along the [011] direction.
  The energy is shown for the proton moving along the center of the channel (impact parameter of 1.98 \AA, red long-dashed line) and along the off-center 
channeling trajectory (impact parameter of 0.7 \AA, blue line). 
  The linear fitting is shown for each case. }
\end{figure}

  The ESP is calculated as $S=dE_{\mathrm{tot}}/dz$, the 
derivative of the Kohn-Sham total 
electronic energy with respect to projectile position $z$ 
along the constant-velocity path~\cite{PhysRevLett.99.235501}.
  $E_{\mathrm{tot}}(t)$ is taken as approximate density-functional for 
$\langle H \rangle (t)$. 
  The average of $S$ is obtained from the slope
of a linear fitting to $E_{\mathrm{tot}}(z)$ as shown in figure~\ref{fitting1}.
  Figure~\ref{fitting1} shows the energy for the case of a proton 
moving in Ge with a velocity of 0.5 a.u. along 
 the center of the [011] channel 
and for an off-center trajectory 
along the same direction.
  The oscillations in the figure reflect the periodicity of
the crystal.
  They are quite prominent for the low-impact factor case, 
  but are not noticeable for the mid-channel trajectory, along which the density is quite homogeneous.

  All the results in the following are 
compared with the SRIM~\cite{srim} data for the 
ESP. 
  SRIM results are obtained semiempirically by averaging over a number 
of different incident directions with distinct impact parameters, 
with no explicit consideration for the channeling conditions studied 
in our work.
  The density of the target materials that we used in the SRIM 
calculations were computed from the unit-cell volume obtained in the 
DFT ground-state calculations.


\section{Results and discussion}


\subsection{Monte Carlo particle transport simulations}\label{sec:res-MC}

\begin{figure}[h!]
\includegraphics[width=0.95\textwidth]{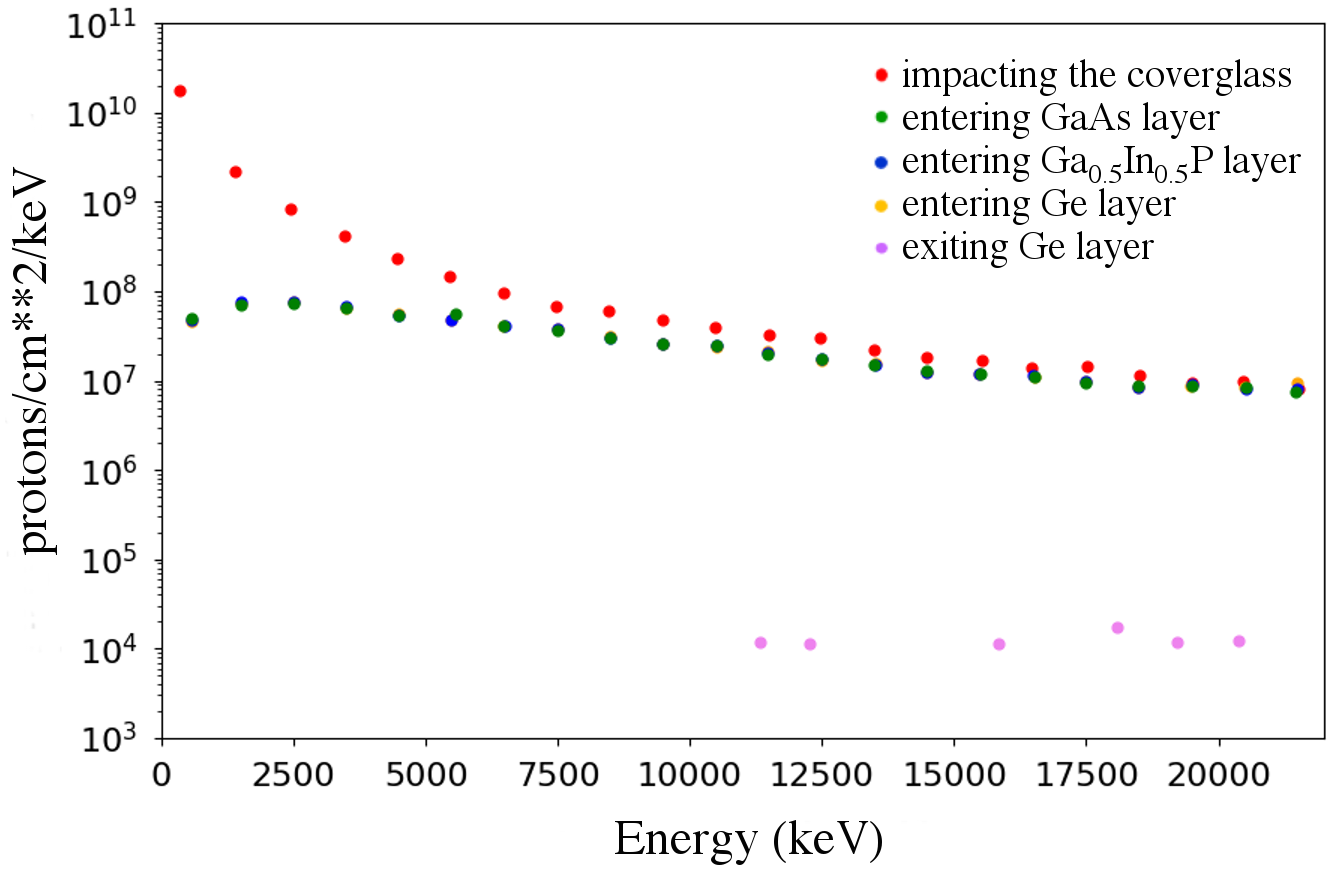}
\caption{\label{Figtrapp_prot} 
 Fluence of the omnidirectional
primary proton radiation across all layers of the solar cell structure (coverglass included).}
\end{figure}

  Figure~\ref{Figtrapp_prot} depicts the results of the 1D-Monte 
Carlo (Geant4-based) simulations of the propagation of the isotropic 
protons through the solar cell stack. The simulations are performed for the representative example case of the protons trapped in the Van Allen belts accumulated during a 3-year 
International Space Station-like mission.
  The first step is the calculation of the slowed-down proton differential 
spectrum through the slab of the protective coverglass SiO$_2$ layer.
  The results in figure~\ref{Figtrapp_prot} clearly show that the coverglass lowers the fluences of 
lower energy particles, leaving almost unaffected the high energy 
portion of the proton spectrum. 
  Those protons are not only the primary impacting protons, but also 
those generated by the nuclear interactions in the target.
  Overall, the results show that, for a realistic scenario,
once the particles are slowed down by the 100 $\mu$m SiO$_2$ coverglass, 
the energies of the particles entering all the other layers change to a 
negligible amount (except for the particles exiting the bottom of the 
bottom Ge subcell, whose spectrum is considerably "hardened", i.e., 
energies lower than $\sim$12 MeV do not exit the whole stack 
(figure~\ref{Figtrapp_prot}).
  Within the protective amorphous oxide cap layer, ions will be 
scattered with no channeling, and only a few will find the channels.
  Contrarily, in the crystalline materials constituting the other 
layers, channeling conditions may either enhance the depth of 
penetration or exhibit, for specific impact parameters, blocking 
trajectories. 
  The fluence always contains particles extending to the low energy 
regime of one-hundredth to tenths of keV, for essentially all layers 
in the structure. 

\begin{figure}[h!]
\includegraphics[width=0.99\textwidth]{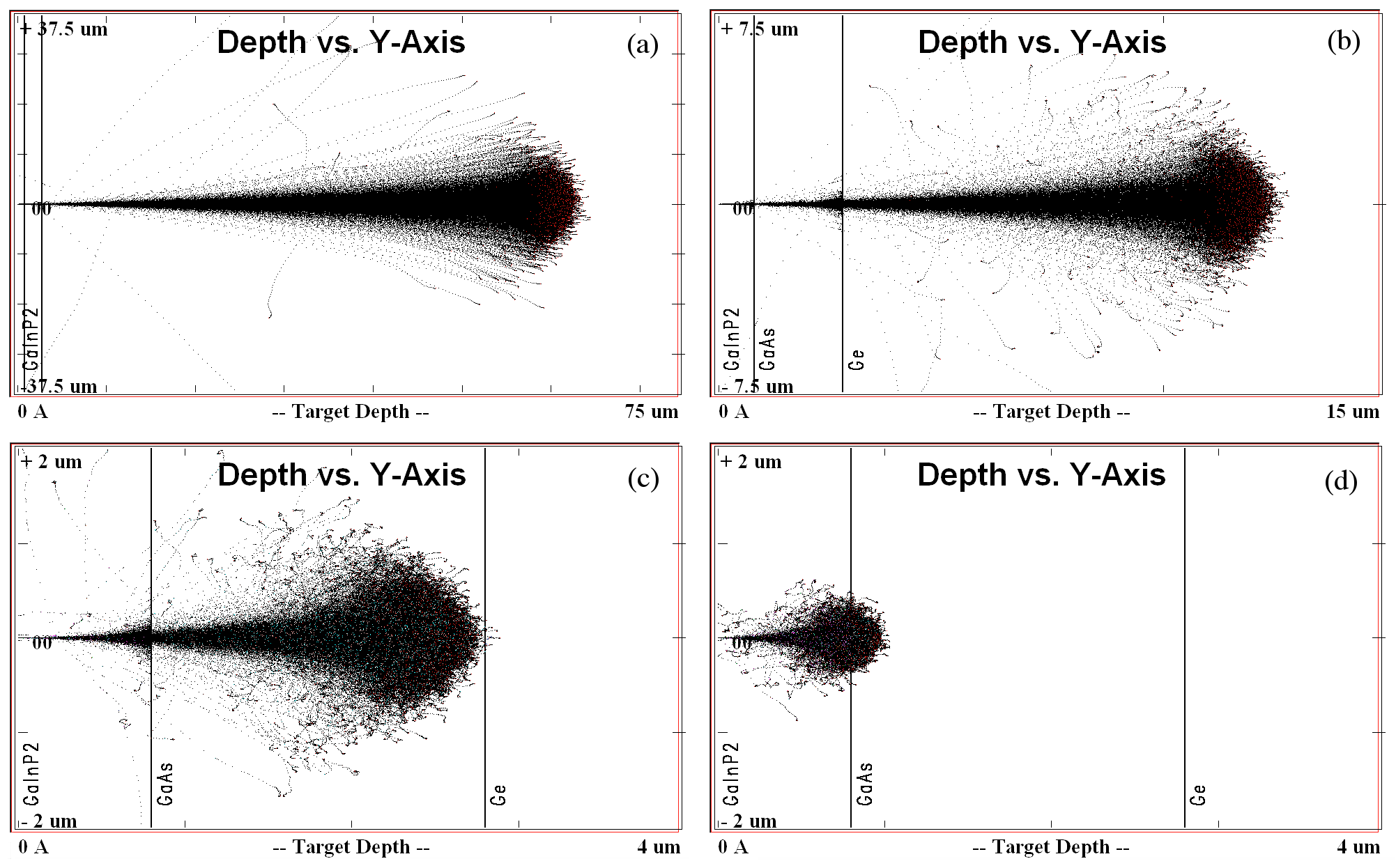}
\caption{\label{Fig-trim} Ion track trajectory output from TRIM for unidirectional monoenergetic protons. 
The thickness of the layers (density) is 0.8 $\mu$m (4.21195 g/cm$^3$) for Ga$_{0.5}$In$_{0.5}$P, 2 $\mu$m (5.8155 g/cm$^3$) for GaAs, and 175 $\mu$m (5.35 g/cm$^3$) for Ge.
The energies of the impacting protons are representative of the low energy ranges used by recent studies: (a) 3 MeV~\cite{AIERKEN201836}, (b) 1 MeV~\cite{KHO2017,DUX2018conf,Donne2019}, (c) 0.3 MeV~\cite{SALS2017}, and (d) 0.1 MeV~\cite{Walt2004}. Note different scale on both axes in different panels.}
\end{figure}

In figure~\ref{Fig-trim}, we report the results of the SRIM (TRIM) Monte Carlo simulations, which are performed, as it is the case in ground testing studies, assuming unshielded configurations and unidirectional (normal incidence), monoenergetic proton irradiation. 
TRIM only deals with Coulomb scattering.
Thus, the calculated damage is that produced by the projectile itself, the primary knock-on atoms (PKAs), and consequently generated secondary (SKAs), and not by eventual ``secondary particles" produced  in nuclear reactions. The energies chosen for the impacting protons are representative of the low energy ranges used by recent studies (3 MeV~\cite{AIERKEN201836}, 1 MeV~\cite{KHO2017,DUX2018conf,Donne2019}, 0.3 MeV~\cite{SALS2017}, and 0.1 MeV~\cite{Walt2004}).
It is important to note that, for such impacting energies, the particles will stop inside the solar cell, in the Ge layer (figure~\ref{Fig-trim}$a,b$), in the GaAs layer (figure~\ref{Fig-trim}$c$), or between the two active layers Ga$_{0.5}$In$_{0.5}$P/GaAs (figure~\ref{Fig-trim}$d$). Thus, the relevant range of impact energies that is necessary to consider in the ESP calculations extends down to 0 keV.

Since the damage produced per unit path length increases as the proton energy decreases and some protons stop within the active region of the cell (figure~\ref{Fig-trim}), a non-uniform damage can be induced at such low energies. Thus, the application of the current modeling approach, the so-called DDD (displacement damage dose) model exhibits some limitations for a relevant energy range (1-10 MeV), since it is based on a uniformity of the damage~\cite{MESS200653} and on an observed linear dependence between the NIEL (i.e., the energy deposited to the atomic displacements) and the damage (i.e., the number of defects produced). Thus the NIEL concept should be improved in order to take into account the details of the structure of the system and the dependence of the NIEL on depth (instead of the NIEL based on the incident energy~\cite{Walt2004}), which can change in channeling vs non-channeling conditions).
If an ion enters a crystalline target along a direction with a low Miller index, the energy-dependent ratio of electronic and nuclear stopping, which is valid for an amorphous material, will no longer be valid.

\subsection{Electronic stopping power in channeling conditions, comparison to SRIM}\label{sec:res-stop-SRIM}

\begin{figure}[ht!]
\centering
\includegraphics[width=0.99\textwidth]{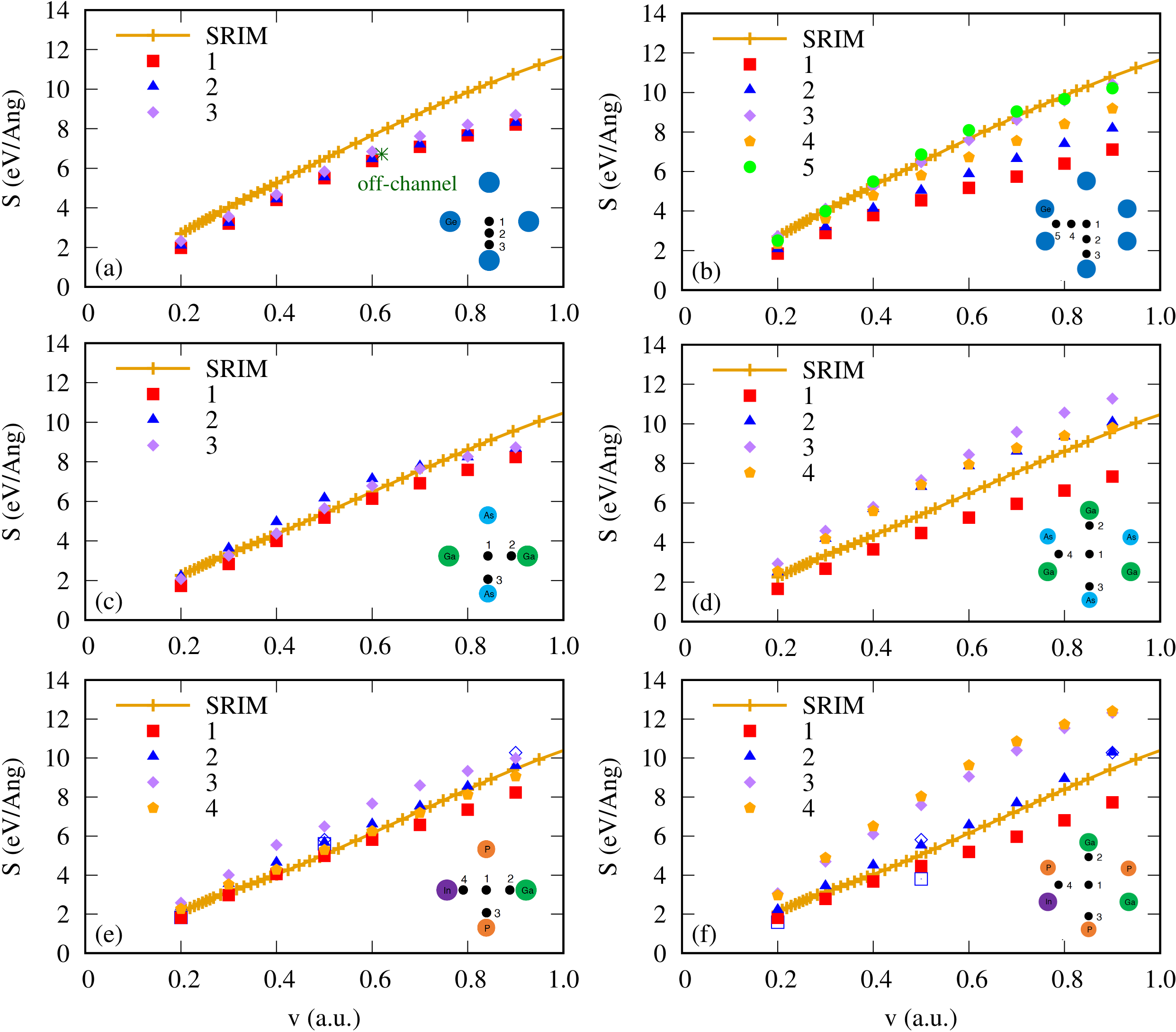}
\caption{ Electronic stopping power for a proton in Ga$_{0.5}$In$_{0.5}$P/GaAs/Ge compared to SRIM. The stopping power is shown as a function of velocity for the channel [001] in (a) Ge, (c) GaAs, and (e) Ga$_{0.5}$In$_{0.5}$P; and for the channel [011] in (b) Ge, (d) GaAs, and (f) Ga$_{0.5}$In$_{0.5}$P. SRIM results are shown as the solid curve with crosses for each layer. A schematic representation of the trajectories of the proton in each channel are shown in the insets. Green asterisk in (a) shows our result for an off-channel trajectory (see text for details). Empty symbols in (e) and (f) are the results from Lee \emph{et al.}~\cite{Lee2018}, the squares and diamonds correspond to the channeling and off-channeling results, respectively (see text for details).}
\label{Figstop-all-SRIM}
\end{figure}

The ESP as a function of the proton velocity is shown in figure~\ref{Figstop-all-SRIM} for three layers of the TJ solar cell and for two channels: [001] and [011]. The corresponding incident energies of the proton can be obtained from $E=m_\mathrm{p} v^2/2$, where $m_\mathrm{p} = 1836$ a.u. is the proton mass (0.999$-$20.234 keV for the chosen range of velocities of 0.2$-$0.9 a.u.). The trajectories of the proton in the [001] and [011] channels are schematically depicted in the insets of figure~\ref{Figstop-all-SRIM}. Trajectory 1 corresponds to the center of the channel in all the cases. The impact parameter for the off-center trajectories closest to the rows of atoms (and to the axis between two atoms in the case of the channel [011]) is 0.7 {\AA} (approximately half the distance from the center of the [001] channel to the target atom). In Ge, we also consider additional trajectories with the impact parameter equal to 1.05 {\AA} for trajectory 2 in figure~\ref{Figstop-all-SRIM}$a$ and 1.4 {\AA} and 1.86 {\AA} for trajectories 2 and 4, respectively, in figure~\ref{Figstop-all-SRIM}$b$.
The orange curve in each panel of figure~\ref{Figstop-all-SRIM} shows the ESP calculated with SRIM~\cite{srim} using the following values of the density: 5.53 g/cm$^3$ for Ge, 5.32 g/cm$^3$ for GaAs, and 5.52 g/cm$^3$ for Ga$_{0.5}$In$_{0.5}$P, which correspond to the lattice constants used in RT-TDDFT calculations.

\begin{figure}[h!]
\centering
\includegraphics[width=0.99\textwidth]{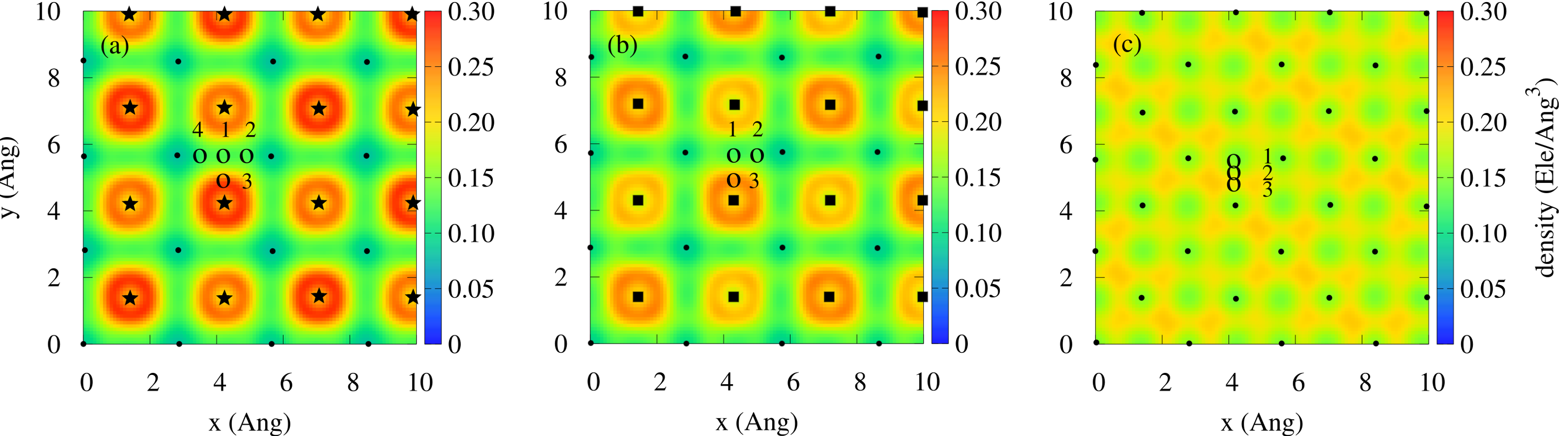}
\caption{Electron density averaged along the [001] channel as a function of position in the perpendicular plane for Ga$_{0.5}$In$_{0.5}$P, GaAs, and Ge. The trajectories of the proton in all systems are indicated with open circles and are the same as in the insets of figure~\ref{Figstop-all-SRIM}. (a) average density in Ga$_{0.5}$In$_{0.5}$P, black dots indicate positions of the alternating Ga/In atoms, while black stars indicate positions of the P atoms; (b) average density in GaAs, black dots and squares show positions of the Ga and As atoms, respectively; (c) average density in Ge, black dots show positions of the Ge atoms. The colour scale is the same for all three panels.}
\label{dens-aver}
\end{figure}

In Ge, the values of the ESP are very similar for all three [001] trajectories (figure~\ref{Figstop-all-SRIM}$a$), since the electronic density distribution does not vary much inside this narrow channel. This can be seen in figure~\ref{dens-aver} in which we plot the electron density averaged along the [001] channel as a function of position in the perpendicular plane for all three layers. The average is obtained from the density calculated for 160 $[x,y]$ planes along the $z$-axis with a step of $dz=0.1$~{\AA}. Figure~\ref{dens-aver}$c$ shows that the average density is much more homogeneous throughout the unit cell in Ge as compared with the other two layers (figure~\ref{dens-aver}$b,c$) and that the density varies only slightly along three trajectories. 

A larger dispersion of ESP is observed in the [011] channel in Ge (figure~\ref{Figstop-all-SRIM}$b$), being a consequence of a larger variation of the electron density inside this channel~\cite{PhysRevB.91.125203}. The lowest ESP comes from the mid-channel, as it can be reasonably expected given the lower electron density encountered in such trajectory. For a slow proton moving along the [011] channel, the stopping power is similar for all the trajectories. However, at higher velocities, the stopping power depends on the impact parameter quite significantly. As it has been noted in Ref.~\cite{PhysRevB.91.125203}, this behavior is a consequence of the strong correlation between the stopping power and the average local density within a small radius of the impact parameter, and that this radius is larger for a slower projectile. 

On average, our results for the ESP in Ge are slightly lower than the SRIM for the proton moving in both channels, except for the trajectories 3 and 5 in the [011] channel, for which the ESP approaches the SRIM values. This suggests that in the SRIM calculations, the average density is higher than the density on most trajectories in the RT-TDDFT calculations for Ge. The average valence density in Ge is about 0.179 electrons/{\AA}$^3$, which corresponds to the trajectory 3 in figure~\ref{dens-aver}$c$, while the average density along the trajectories 1 and 2 is slightly lower. Present results for a proton moving on the mid-channel trajectories along the directions [001] and [011] in Ge are in agreement with the results previously obtained by Ullah \emph{et al.}~\cite{PhysRevB.91.125203}.

In figure~\ref{Figstop-all-SRIM}$a$, we show as well the ESP for an off-channel proton trajectory (random direction, not parallel to any channel) in order to see how well it compares to SRIM. The random trajectory is arbitrarily chosen so that it makes 14$^\circ$ with the direction [001] ($z-$axis). The initial projectile position is at the center of the [001] channel at a distance 0 {\AA} from the first atomic plane perpendicular to this direction. The velocity of the proton is 0.62 a.u. (0.6 a.u. along the $z-$axis and 0.15 a.u. along the $x-$axis). The calculation was performed during 6 fs, four times the time needed to cross the simulation box at this velocity. Thus it is equivalent to four different random trajectories~\cite{Lee2018}. The resulting ESP for the off-channel trajectory is in between the values for two off-center channeling trajectories 2 and 3, and thus does not bring the channeling results closer to SRIM. It is known that longer trajectories would affect our results, but on a smaller scale
than the observed discrepancy~\cite{Lee2018}. A possible reason for that is the absence of core electrons, as discussed in the next subsection. However,
SRIM is expected to be inaccurate on that scale as well~\cite{Paul2013}, and further expense in calculations is therefore not justified.

In GaAs, similarly to Ge, the ESP for the proton in the [001] channel is almost independent on the impact parameter and agrees perfectly with the SRIM data (figure~\ref{Figstop-all-SRIM}$c$). Only the ESP for the proton on trajectory 3 is slightly higher than on 1 and 2, given the higher average electron density on this trajectory (figure~\ref{dens-aver}$b$).
In the [011] channel, the ESP is below SRIM for the proton moving on the mid-channel trajectory, while it is above SRIM for all the off-center trajectories with small impact parameter (i.e., higher electron density) (figure~\ref{Figstop-all-SRIM}$d$). At proton velocities below 0.5 a.u., the ESP in the [011] channel is similar for all the off-center trajectories. At higher velocities, however, 
the ESP corresponding to the trajectory 3, closest to the As atoms, is slightly higher. This could indicate that the maximum of the ESP is located at different velocities for different proton trajectories.

In the case of the Ga$_{0.5}$In$_{0.5}$P layer, again, on average, our results are in good agreement with SRIM (figure~\ref{Figstop-all-SRIM}$e,f$). However, the dependence of the ESP on the trajectory is observed in both channels. For the proton moving on trajectories close to Ga and In atoms in both channels, the ESP values are very close, owing to the fact that both elements have three electrons in the outer shell. A much higher ESP is obtained for the trajectories close to P, which have five valence electrons, and thus, the electron density is higher along these trajectories (trajectory 3 in figure~\ref{dens-aver}$a$). 
Analysis of the average electron density for different proton trajectories in the direction [001] in Ga$_{0.5}$In$_{0.5}$P, indicated with open circles in figure~\ref{dens-aver}$a$, shows that the density is similar for trajectories 1, 2, and 4, but is higher by a factor of 2 along the trajectory 3. However, the stopping power for a proton moving on the trajectory 4 is only 1.4 (maximum, at $v=0.7$ a.u.) times larger than the one on the trajectory 1 (figure~\ref{Figstop-all-SRIM}$e$). Thus, the stopping power is not linear with the average density along the proton path.

In figure~\ref{Figstop-all-SRIM}$e,f$ we also compare our results to the calculations by Lee \emph{et al.}~\cite{Lee2018}. Similarly to our work, the ESP is obtained with RT-TDDFT (LDA), but using a plane-wave basis set. They are represented as empty squares for the stopping along the mid-channel trajectory, and as empty diamonds for the off-channel trajectory (i.e., random trajectory through the lattice, not parallel to any of the main crystallographic directions). There are small deviations between the results of both approaches, probably due to basis sets and pseudopotentials, but the main results, trends and conclusions are the same. Those deviations are smaller than the spread for different trajectories, and the differences with respect to SRIM.

Small deviation of the ESP from the SRIM behavior observed at high velocities (especially in Ge) is due to two reasons. First, closer to the maximum of the stopping power, the effect of the inner electrons starts to be important, which is not taken into account in these calculations. We will discuss this effect in the next subsection. Another reason is that the Sankey integrator used in SIESTA to propagate the KS states in time~\cite{PhysRevB.95.115155} is known to be problematic at higher energies.


\subsection{Comparative analysis of the ESP in different layers}\label{sec:res-stop}

\begin{figure}[ht]
\centering
\includegraphics[width=0.99\textwidth]{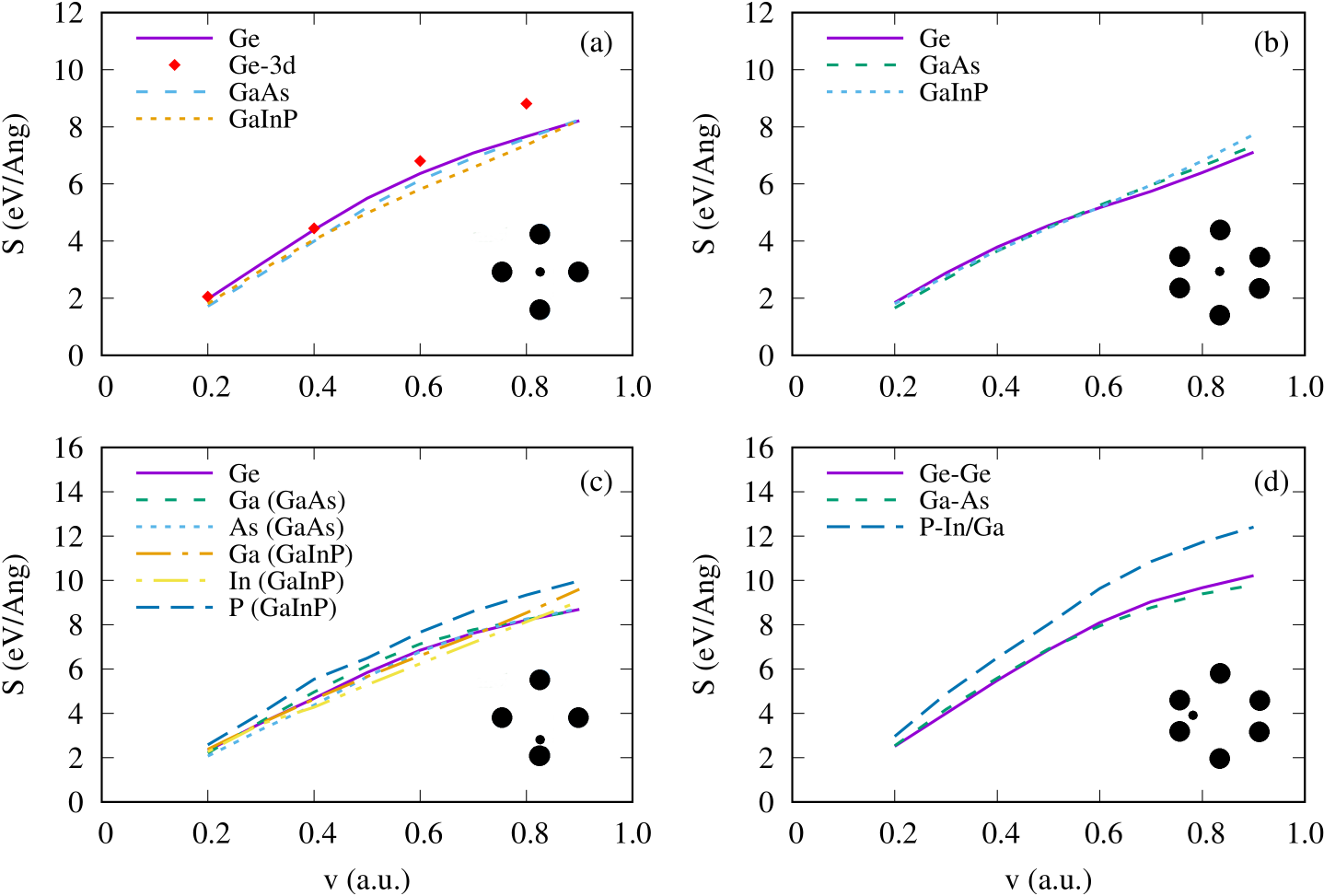}
\caption{Electronic stopping power for a proton in Ga$_{0.5}$In$_{0.5}$P/GaAs/Ge, a comparison of different materials. The stopping power is shown as a function of velocity for (a) direction [001], center of the channel. Red diamonds show the results for Ge including semicore electrons; (b) direction [011], center of the channel; (c) direction [001], impact parameter of 0.7 \AA; (d) direction [011], impact parameter of 0.7 \AA~from the edge of the hexagon. The inset on each panel shows a schematic view of the channel section and the proton position according to the impact parameter.}
\label{Figstop-all}
\end{figure}

  We further analyze our results by comparing the values
of the ESP for equivalent proton trajectories inside different layers of the TJ solar cell as shown in figure~\ref{Figstop-all}.

Overall, the ESP do not vary much from material to material for the proton moving on the mid-center trajectories (figure~\ref{Figstop-all}$a,b$). For the smallest impact parameter of 0.7 {\AA} in the [001] channel (figure~\ref{Figstop-all}$c$), we show the ESP for a proton trajectories close to different elements in each material. The highest ESP corresponds to the trajectory close to P atoms in Ga$_{0.5}$In$_{0.5}$P. Figure~\ref{Figstop-all}$d$ shows the ESP for the proton trajectories 5 (Ge) and 4 (GaAs and Ga$_{0.5}$In$_{0.5}$P) in the [011] channel (see trajectories in figure~\ref{Figstop-all-SRIM}). Here as well, the largest ESP is obtained for the proton on the trajectory close to P atoms. 

The importance of core electrons was checked for the case of Ge (figure~\ref{Figstop-all}$a$) by including 3d electrons into the valence shell (see Appendix for details of the calculations). At velocities below 0.6 a.u., the addition of core electrons has no effect on the ESP. However, at velocity 0.8 a.u, the ESP is notably higher when 3d electrons are included in the valence, for this particular trajectory. 

\begin{figure}[ht]
\centering
\includegraphics[width=0.99\textwidth]{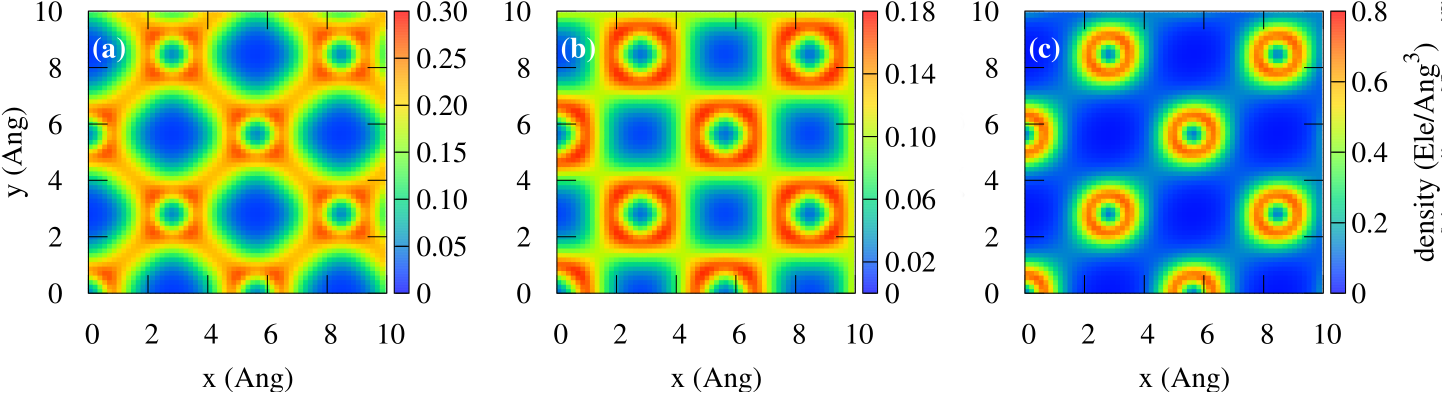}
\caption{Electron density for three planes in Ga$_{0.5}$In$_{0.5}$P perpendicular to the direction [001]. The planes are crossing through (a) Ga; (b) In; (c) P atoms.}
\label{dens}
\end{figure}

Since Ga$_{0.5}$In$_{0.5}$P is a compound with the largest number of constituent species in TJ solar cells, it is interesting to look at the electron density distribution inside it and try to correlate it with the observed ESP. Figure~\ref{dens} shows the electron density for three different planes perpendicular to the [001] direction. Each plane crosses the positions of atoms of only one type (Ga, In, or P). From figure~\ref{dens}, it is obvious that the electron density in the vicinity of the P atoms is much higher, which explains the largest ESP for protons on the trajectories 3 and 4 (figure~\ref{Figstop-all-SRIM}). 

\begin{figure}[h!]
\centering
\includegraphics[width=0.7\textwidth]{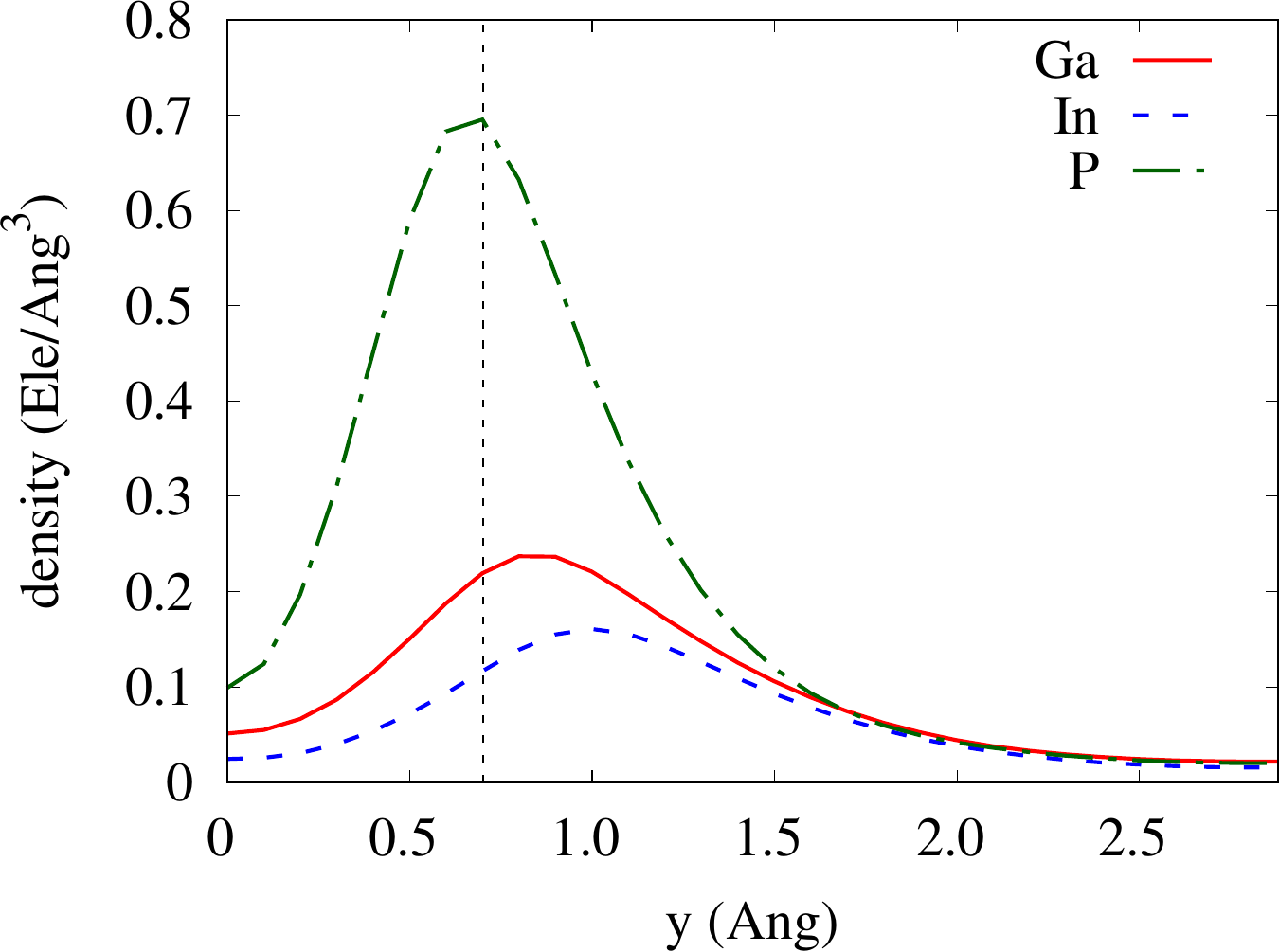}
\caption{Electron density along the $y$-axis, corresponding to the cut through $x=0$ in three panels of figure~\ref{dens}. The vertical dashed line shows the position of the proton in the off-center channeling trajectory at 0.7 {\AA} from the atomic position, corresponding to trajectories 2, 3, and 4 in the inset of figure~\ref{Figstop-all-SRIM}$e$.}
\label{dens-x-0}
\end{figure}

Not only the value, but the spatial distribution of the electron density varies in different planes as well. Figure~\ref{dens-x-0} shows the density as a function of $y$ coordinate for all three panels of figure~\ref{dens} corresponding to $x=0$. Thus three curves in figure~\ref{dens-x-0} show the density distribution around one single atom of Ga, In, and P, respectively. The system of coordinates is chosen so that all the atoms are centered at the origin ($y=0$). The density, going down towards the nuclear position, reflects the absence of core electrons in the calculations. The maximum density around P atom is higher and is closer to the origin. Thus the projectile moving at 0.7 {\AA} from the atomic position (indicated by the vertical dashed line in figure~\ref{dens-x-0}), in the case of P, is moving in the area of much higher electron density than in the case of Ga and In.

\subsection{Comparison with the homogeneous electron gas model}

\begin{figure}[ht]
\centering
\includegraphics[width=0.7\textwidth]{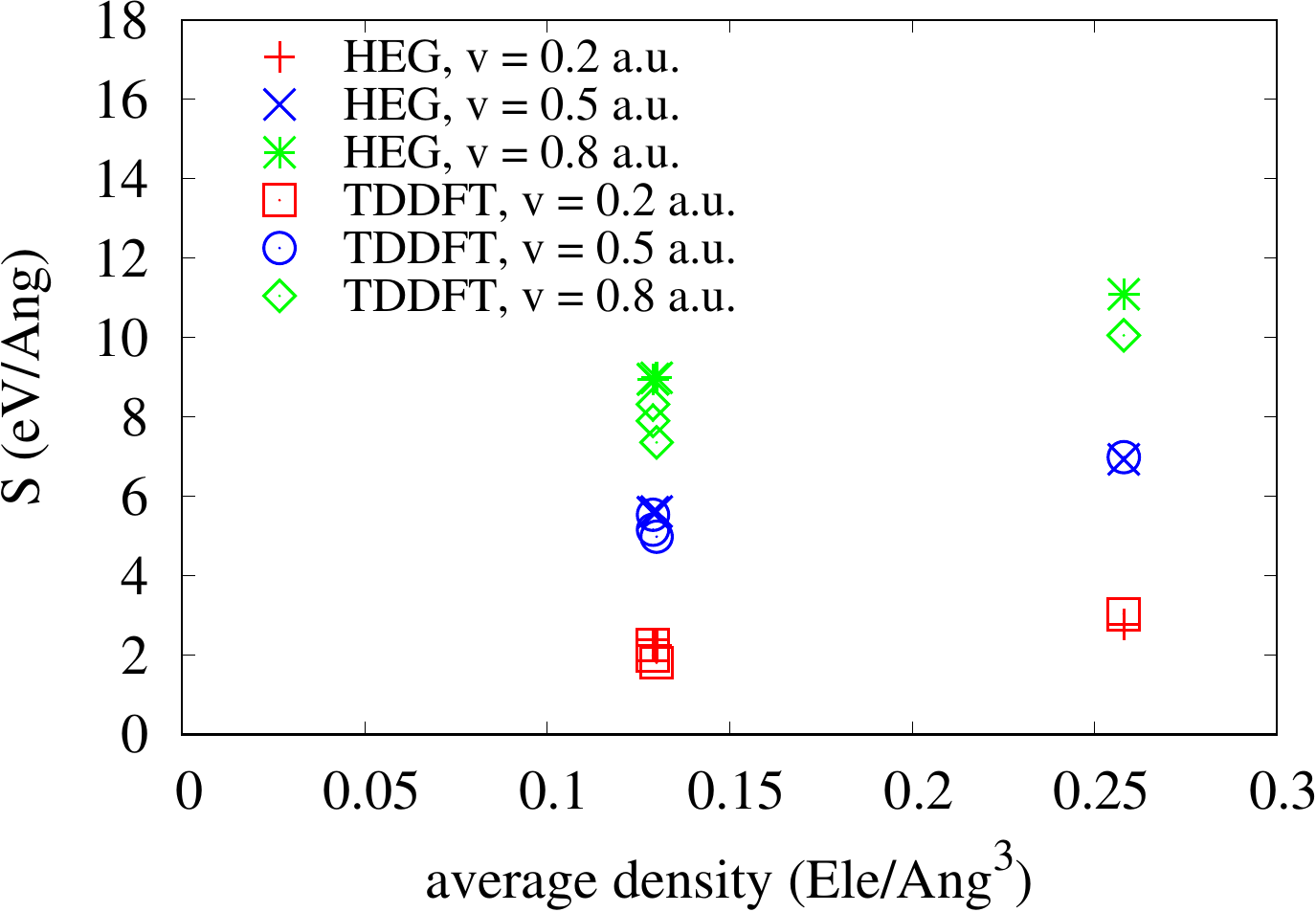}
\caption{Electronic stopping power from RT-TDDFT compared to the HEG approximation. The stopping power as a function of the average electron density in Ga$_{0.5}$In$_{0.5}$P channel [001] from RT-TDDFT for velocities of the proton $v=0.2$ a.u. (open squares),  $v=0.5$ a.u. (open circles), and $v=0.8$ a.u. (open diamonds); and HEG approximation for velocities $v=0.2$ a.u. (pluses),  $v=0.5$ a.u. (crosses), and $v=0.8$ a.u. (asterisks).}
\label{s-HEG}
\end{figure}

In order to show the non-linearity of the ESP obtained with RT-TDDFT, we compare our results with the homogeneous electron gas (HEG) approximation. In HEG, the electron density is defined through the so-called Wigner-Seitz radius $r_{\mathrm{s}}$, a one-electron radius, as $n=3/(4 \pi r_{\mathrm{s}}^3)$. The stopping power in HEG is calculated as a product of the friction coefficient and proton velocity. The friction coefficients corresponding to different $r_{\mathrm{s}}$ are taken from~\cite{bib:juaristi08}. In figure~\ref{s-HEG} we show the stopping power as a function of the average electron density in Ga$_{0.5}$In$_{0.5}$P channel [001]. The lower density points (three points close to each other) correspond to the trajectories 1, 2, and 3, and the higher density corresponds to the trajectory 4. The crosses show the results of the HEG approximation, while the squares show the RT-TDDFT stopping power. The comparison is shown for three different velocities of the proton. At lower velocities, the agreement is very good. At higher velocity, however, the importance of the non-linear effects is clearly observed as the HEG results overestimate our results.

\subsection{Interface effects on the proton energy loss}\label{sec:res-interf}

\begin{figure}[ht]
\centering
\includegraphics[width=0.9\textwidth]{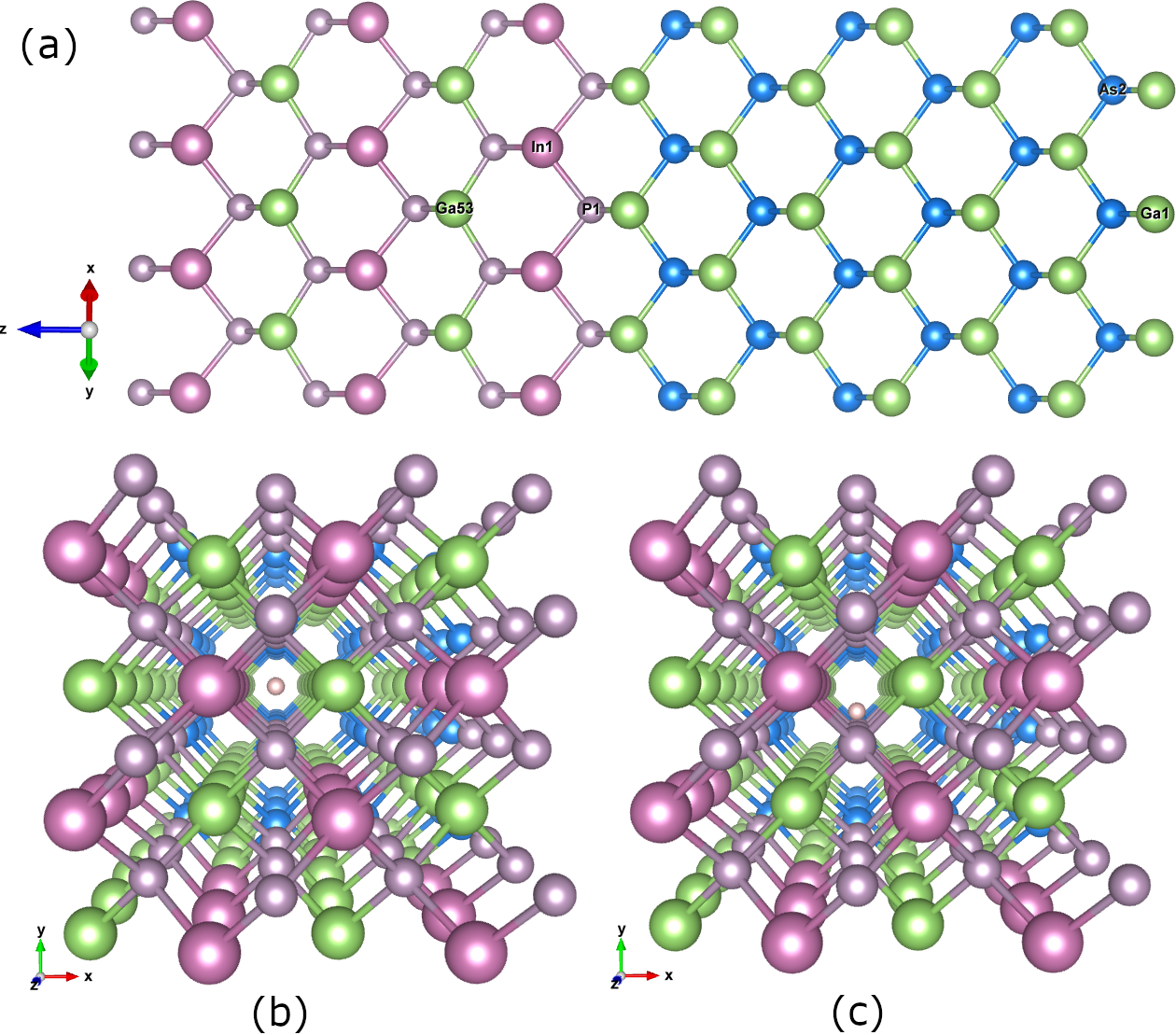}
\caption{ Atomic structure of the interface. (a) Structure of the interface between Ga$_{0.5}$In$_{0.5}$P and GaAs, (b) proton in the center of the channel [001], (c) proton in the off-center channel at 0.7 {\AA} from P and As atoms of the interface structure.}
\label{Figinterface}
\end{figure}

In order to study the effect of an interface on the energy loss of the proton, we constructed a super-cell of the interface between the two upper layers of the TJ solar cell, i.e., Ga$_{0.5}$In$_{0.5}$P/GaAs along the crystallographic direction [001] of the epitaxial growth of the lattice-matched solar cells (figure~\ref{Figinterface}$a$). The super-cell consists of 192 atoms and has the lattice parameter average between the pure bulk Ga$_{0.5}$In$_{0.5}$P and GaAs. We have chosen two trajectories of the proton through the interface, the channeling one (figure~\ref{Figinterface}$b$) and the off-center channeling trajectory (figure~\ref{Figinterface}$c$), in which the proton first passes close to P and then to As atomic rows in the direction [001] with an impact parameter of 0.7 {\AA}.

\begin{figure}[h!]
\centering
\includegraphics[width=0.99\textwidth]{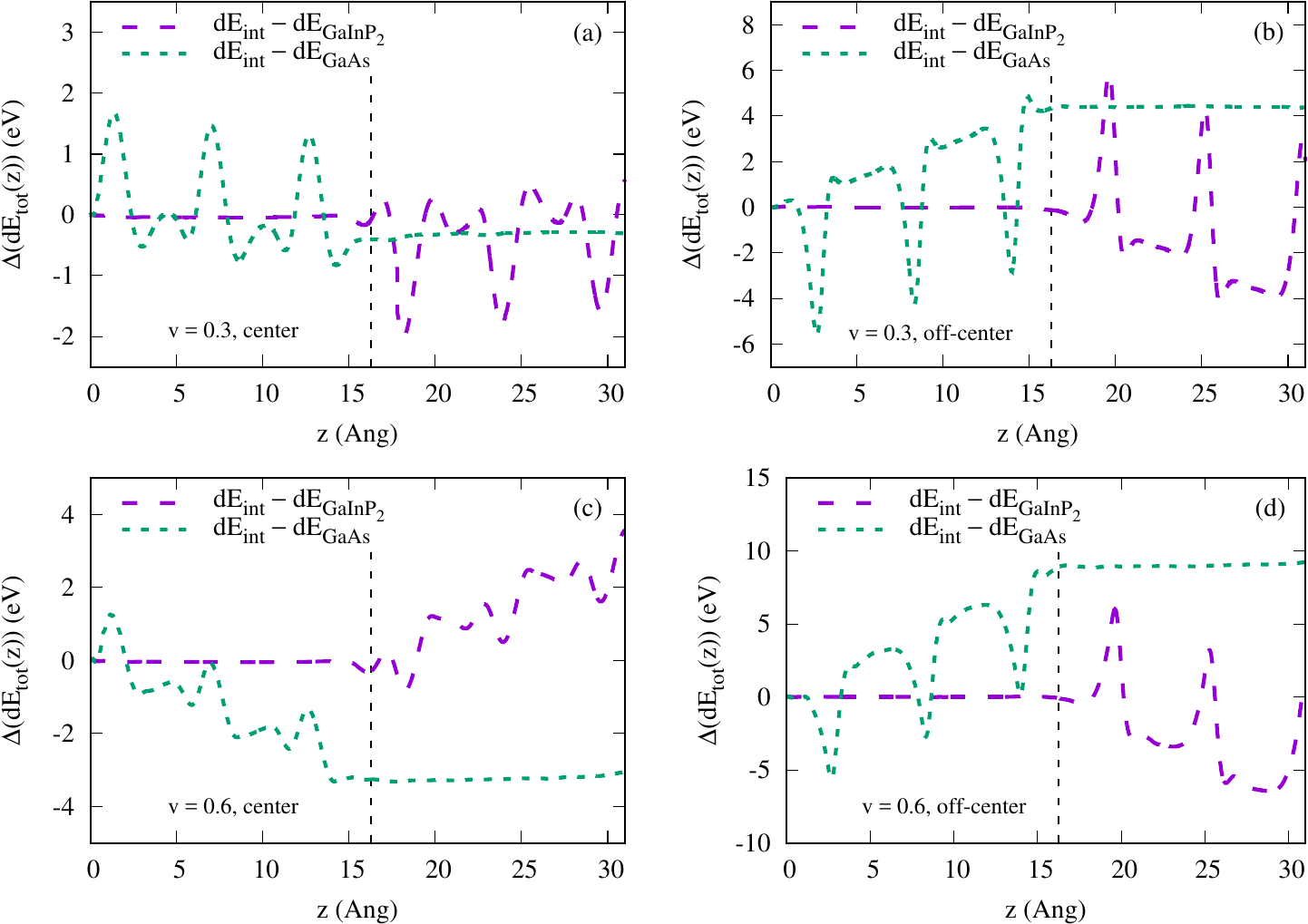}
\caption{Difference in the proton energy loss between pure GaAs and Ga$_{0.5}$In$_{0.5}$P and the interface between the two materials. The quantity plotted is $\Delta(dE_{\mathrm{tot}}(z))=dE_{\mathrm{int}}(z)-dE_{\mathrm{pure}}(z)$, where $dE(z)=E(z)-E(z=0)$ for (a) channeling trajectory, $v=0.3$ a.u.; (b) impact parameter 0.7 {\AA}, $v=0.3$ a.u.; (c) channeling trajectory, $v=0.6$ a.u.; (d) impact parameter 0.7 {\AA}, $v=0.6$ a.u., as indicated in each panel. The interface between GaAs and Ga$_{0.5}$In$_{0.5}$P is located at 16.26 {\AA} and is showed by the vertical dashed line in each panel.}
\label{Figint-delta_dE}
\end{figure}

The difference in the energy loss of the proton moving through the interface and through the pure Ga$_{0.5}$In$_{0.5}$P and GaAs super-cells of the same size is shown in figure~\ref{Figint-delta_dE} for velocities 0.3 a.u. and 0.6 a.u. for both trajectories. The interface between two materials is located at $z_{\mathrm{int}} = 16.26$ {\AA}. The proton first moves though the Ga$_{0.5}$In$_{0.5}$P, hence the difference in the proton energy loss $\Delta(dE_{\mathrm{tot}}(z))=dE_{\mathrm{int}}(z)-dE_{\mathrm{GaInP_2}}(z)$ is equal to zero until $z \simeq z_{\mathrm{int}}$ in all the cases. For a similar reason, starting from $z \simeq z_{\mathrm{int}}$, the value of $\Delta(dE_{\mathrm{tot}}(z))=dE_{\mathrm{int}}(z)-dE_{\mathrm{GaAs}}(z)$ is constant. Interestingly, the slope of $\Delta(dE_{\mathrm{tot}}(z))$ has opposite trends for the two trajectories. In the case of the mid-channel trajectory (figure~\ref{Figint-delta_dE}$a,c$), the proton energy loss in GaAs is larger than in Ga$_{0.5}$In$_{0.5}$P. On the off-center trajectory (figure~\ref{Figint-delta_dE}$b,d$), however, the behavior is reversed. This change of slope occurs due to the fact that the electron density around P atoms is more localized (see figure~\ref{dens}), and is slightly lower than around Ga in the center of the [001] channel ($y=2.1$ {\AA}, figure~\ref{dens-x-0}). Thus, the electron density in the center of the channel in Ga$_{0.5}$In$_{0.5}$P is expected to be lower than in GaAs, giving rise to lower energy loss of the proton moving on this trajectory in Ga$_{0.5}$In$_{0.5}$P. On the off-center trajectory, on the contrary, the density around P is much higher. This behaviour is, however, a bulk effect not related to the interface itself.

To quantify the interface effect and separate it from the bulk effects, we have calculated the proton energy loss on a given part of the trajectory equal to the length of two unit cells on each side of the interface structure ($dE_{\mathrm{int(Ga_{0.5}In_{0.5}P)}}$ and $dE_{\mathrm{int(GaAs)}}$) and for the equivalent part of the trajectory of pure structures ($dE_{\mathrm{Ga_{0.5}In_{0.5}P}}$ and $dE_{\mathrm{GaAs}}$). The difference between the energy loss of a proton moving through the interface structure and the sum of the energy losses of a proton moving through the corresponding intervals of pure structures, gives us the difference due to the interface: $\Delta E_{\mathrm{int}}=dE_{\mathrm{int(Ga_{0.5}In_{0.5}P)}}+dE_{\mathrm{int(GaAs)}}-dE_{\mathrm{Ga_{0.5}In_{0.5}P}}-dE_{\mathrm{GaAs}}$. The values we obtained for $\Delta E_{\mathrm{int}}$ for center and off-center channeling trajectories are $0.39$ eV and 0.60 eV for velocity 0.3 a.u., and $-0.075$ eV and 0.072 eV for velocity 0.6 a.u., respectively. Thus, the interface effect is higher at lower velocities. The energy loss due to the interface even becomes negative at high velocity in the case of center-channel trajectory, meaning effectively an energy gain. These differences, however, are $1-2$ orders of magnitude smaller than the difference between the materials, i.e., the bulk effect.

%

\subsection{Effect of strain in lattice-matched solar cells on the ESP}\label{sec:res-strain}

Most of the TJ solar cells currently used for space applications are lattice-matched. Ge is the thickest layer used as a substrate. The GaAs layer is grown on top of Ge and the Ga$_{0.5}$In$_{0.5}$P is grown on top of the GaAs layer, with the lattices of different layers matching at the interface. The experimental lattice constants of these materials are very similar, 5.658 {\AA} for Ge, 5.653 {\AA} for GaAs (only $0.08\%$ different from Ge), and 5.6596 {\AA} for Ga$_{0.5}$In$_{0.5}$P. This small difference in the lattice constant leads to a strain in the GaAs and Ga$_{0.5}$In$_{0.5}$P layers. We have analysed the effect of strain in two cases. In the first case, we have calculated the stopping power for a proton moving with $v=0.8$ a.u in GaAs with lattice constants 5.65 and 5.75 {\AA} (the difference of 1.74\%). The stopping power only changed by 0.28\%. In another case, we compared the stopping power for a proton moving with $v=0.5$ a.u in GaAs with lattice constants 5.75 and 5.57 {\AA} (3.13\% different). The difference in the stopping power was only 1.13\%. The actual differences in the lattice constants of the three layers of TJ solar cell are much smaller, and the changes in the stopping power can be considered negligible.

\section{Conclusions}

In this work, using RT-TDDFT we have calculated the electronic stopping power of the three sub-junctions of the TJ solar cells for the impacting protons. Kinetic energies of the proton are considered in the keV range, which we have shown to be relevant in the transport of protons through the full stack in a realistic scenario. 
We compared our results for channeling conditions to the SRIM semi-empirical stopping power. Our results have shown that in Ge, the stopping power in the channel [001] is almost independent on the trajectory, while in both compounds (GaAs and Ga$_{0.5}$In$_{0.5}$P) the electronic stopping varies more along different trajectories, affected by varying electron density. In the case of the channel [011], strong dependence of the stopping power on the trajectory is observed in all three materials. The effect is more significant at higher proton velocities.

The change of sign for the slope of the energy loss difference between GaAs and Ga$_{0.5}$In$_{0.5}$P in the interface structure is a bulk effect, which is explained in terms of electron density. Namely, the average density along the center of the [001] channel in GaAs is higher than in Ga$_{0.5}$In$_{0.5}$P, while it is the other way around along the off-center-channel trajectory.

The effect of the interface between the layers of the lattice-matched multilayer solar cell, as well as the effect of strain, have been found to be negligible, which is understood, given the very similar chemistry and lattice constants of the three materials. The interface effect is found to be higher for lower velocity. At higher velocity, the energy loss can become negative, which effectively means that the proton loses less energy due to the interface.

Additional studies can help to understand the importance of the channeling effect in reducing the damage caused by radiation. The coupling between nuclear and electronic stopping power will be considered as a further step in this study.

\section*{Appendix A. SIESTA calculation parameters}

  All the DFT and RT-TDDFT calculations were performed using
SIESTA version $siesta/4.1b4$-$trunk$-750. The plane-wave cutoff of the finite 3D grid of 200 Ry was used.

The basis functions of the double-$\zeta$ polarized basis set have been generated
as explained in Refs.~\cite{Artacho99,Junquera2001}.
The cutoff radii of the $\zeta_1$ 
function were defined by an energy shift of 272.1 meV (20 meV for Ga$_{0.5}$In$_{0.5}$P and 10 meV in all the interface calculations), and the ones 
for the $\zeta_2$ function, using the split-norm parameter of 
15\% (30\% for Ga$_{0.5}$In$_{0.5}$P and in all the interface calculations). The parameters needed to generate the basis set in the 
calculation with $3d$-shell in Ge are listed in 
Table~\ref{tab_basis}. The extra
$d-$shell is introduced as polarization orbital and has been 
obtained using the electric field polarization scheme described in
Ref.~\cite{Soler2002}, and, therefore, its finite-support radius coincides
with that of the shell it polarizes.
The double-$\zeta$ polarized basis set of the SIESTA code have previously been checked against plane-wave code ABINIT in similar calculations for Ge~\cite{PhysRevB.91.125203}. The agreement between the two codes for the band structure of Ge was excellent. Moreover, the validity of using the double-$\zeta$ polarized basis set in this work is verified by comparing our results for the stopping power of the Ga$_{0.5}$In$_{0.5}$P with the plane-wave results from Ref.~\cite{Lee2018}. A reasonable agreement is observed for equivalent trajectories.

%
\begin{table}[!h]
\centering
\caption{
Cutoff radii $r(\zeta_1)$ and $r(\zeta_2)$ of the first and second $\zeta$ functions, respectively (Bohr); soft-confinement potential prefactor $V_0$ (Ry); inner radius of the soft-confinement potential $r_i$. }
\begin{tabular}{lcccccc}
\hline
Species & $n$ & $l$ & $r(\zeta_1)$ & $r(\zeta_2)$ & $V_0$ & $r_i/r(\zeta_1)$ \\ \hline
 Ge & 3 & 2 & 4.5 & 3.5 & 50 & 0.9 \\ 
    & 4 & 0 & 8.0 & 7.0 & 50 & 0.9 \\
    & 4 & 1 & 8.0 & 4.0 & 50 & 0.9 \\
    & 4 & 3 & 2.5 & 2.0 & 50 & 0.9 \\
    \hline
\end{tabular}
\label{tab_basis}
\end{table}

The pseudopotentials have been generated following Ref.~\cite{PhysRevB.43.1993} with the parameters listed in Table~\ref{tab_pseudo}. For all the element except for H, the pseudopotentials were generated with the partial core correction~\cite{PhysRevB.26.1738}, with the radii $r_{\mathrm{pc}}$ given in Table~\ref{tab_pseudo}.

\begin{table}[!h]
\centering
\caption{
Pseudopotentials cutoff radii for each angular-momentum channel (in Bohrs) and the partial core-correction radii $r_{\mathrm{pc}}$ (Bohr). }
\begin{tabular}{lccccc}
\hline
Species & $s$ & $p$ & $d$ & $f$ & $r_{\mathrm{pc}}$ \\ \hline
 H  ($1s^2$) & 1.25 & 1.25 & 1.25 & 1.25 & \\
 Ge ($4s^24p^2$) & 2.25 & 2.99 & 2.48 & 2.48 & 1.958 \\ 
 Ge ($3d^{10}4s^24p^2$) & 1.49 & 1.96 & 1.96 & 1.96 & 1.958  \\
 Ga ($4s^24p^1$) & 2.18 & 2.35 & 2.53 & 2.53 & 1.389 \\
 As ($4s^24p^3$) & 2.05 & 2.21 & 2.50 & 2.50 & 2.728 \\
 In ($5s^25p^1$) & 2.51 & 2.61 & 2.85 & 2.42 & 1.395 \\
 P  ($3s^23p^3$) & 1.70 & 1.88 & 2.00 & 2.00 & 0.992 \\
    \hline
\end{tabular}
\label{tab_pseudo}
\end{table}

\vskip1pc

\section{Acknowledgements}

The work presented has been fund by the Research Executive Agency under the European Union's Horizon 2020 Research and Innovation program (project ESC2RAD: Enabling Smart Computations to study space RADiation effects, Grant Agreement 776410).
The authors thankfully acknowledge the computer resources at Donostia international Physics Center and at MareNostrum and the technical support provided by Barcelona Supercomputing Center (FI-2019-2-0017).

\pagebreak


\end{document}